# Unraveling the collinearity in short-range order parameters for lattice configurations arising from topological constraints

Abhijit Chatterjee

Department of Chemical Engineering, Indian Institute of Technology Bombay, Mumbai 400076, Maharashtra, India

Email: abhijit@che.iitb.ac.in

**Abstract**

In multicomponent lattice problems, e.g., in alloys, and at crystalline surfaces and interfaces, atomic arrangements exhibit spatial correlations that dictate the kinetic and thermodynamic phase behavior. These correlations emerge from interparticle interactions and are frequently reported in terms of the short-range order (SRO) parameter. Expressed usually in terms of pair distributions and other cluster probabilities, the SRO parameter gives the likelihood of finding atoms/molecules of a particular type in the vicinity of others atoms. This study focuses on fundamental constraints involving the SRO parameters that are imposed by the underlying lattice topology. Using a data-driven approach, we uncover the interrelationships between different SRO parameters (e.g., pairs, triplets, quadruplets, etc.) on a lattice. The main finding is that while some SRO parameters are independent, the remaining are collinear, i.e., the latter are dictated by the independent ones through linear relationships. A kinetic and thermodynamic modeling framework based on these constraints is introduced.

# 1. Introduction

The Metropolis Monte Carlo (MMC) method [1] has been a popular choice for studying molecular arrangements in condensed matter systems. While the applications of MMC are quite widespread, the topic of this paper is the study of configurational disorder in crystalline systems. Solid materials containing two and more components are commonly encountered in fields of catalysis [2–5], ionic conductors [6], alloys [7,8], separations, and many others [9,10]. Although the material has a well-defined crystal structure, it often possesses configurational disorder (see Figure 1a). Depending on the interactions, the local particle arrangement can be anywhere between perfectly-random to perfectly-ordered. The configurational disorder is quantified in terms of short-range order (SRO) parameters, such as the radial distribution function (rdf). Such quantities are used for interpreting atomistic configurations from MMC [11–13], predicting, designing, and optimizing material properties, and even for accelerating MMC [14].

In liquids, historically, several analytical treatments of configurational disorder have been attempted. A hierarchy of correlation functions is obtained by integrating-out phase space coordinates [15]. Solving for the exact two-particle distribution can be elusive, since information about other particle distributions, which are also unknown, is required. To address this problem, physically-motivated closure relations are often invoked. For instance, one may write the three-particle distribution as a product of three two-particle distributions. However, such approximations also introduce inaccuracies. Clearly there is a need to understand the relationships between various $N$-particle distributions. Here we employ a data-driven approach to explore some fundamental properties of SRO parameters. We restrict ourselves to a lattice system, where the distributions are sharply-peaked/discrete in space.

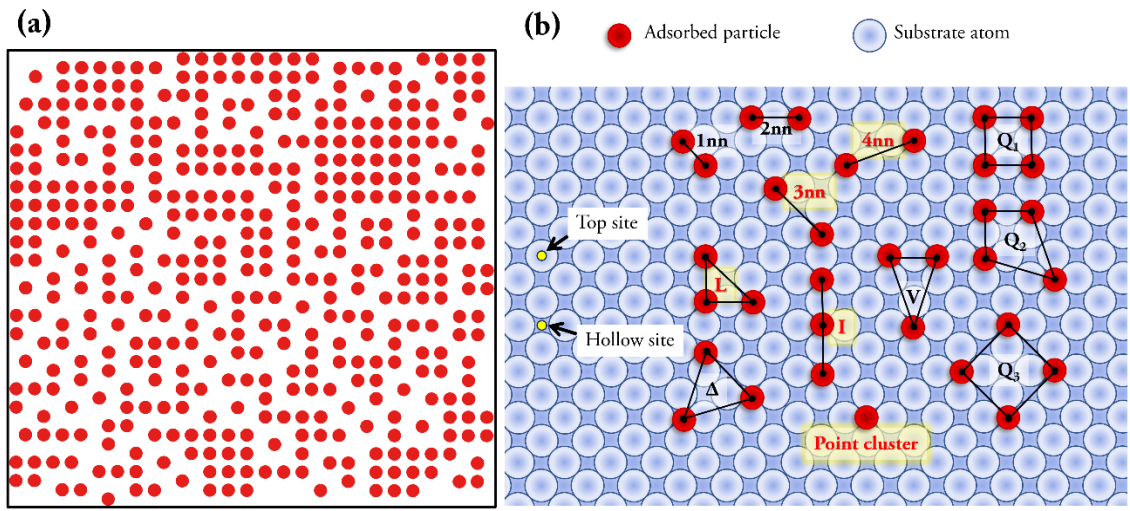

**Figure 1. (a) A lattice configuration with a partially ordered structure. Short-range order (SRO) statistics are based on many-particle clusters. (b) Corresponding detailed schematic showing binding sites and particle clusters. The substrate, here a metal (100) surface, is shown for clarity. Particles are allowed to adsorb only at the hollow sites.**

Consider a binary $A_xB_{1-x}$ lattice system involving a single type of site that can be occupied by either $A$ or $B$. The simplest SRO parameter is the pair probability. Given that a site has an $A$ particle, the probability of finding an $A$ particle at the $n^{th}$ neighbor site is specified. Probabilities involving the $A-B$ and $B-B$ pairs are related to the $A-A$ pair [14,16], and are not discussed. Suppose it is stated that the first nearest neighbor (1nn) pairs are absent – this conveys a lot about the possible configurations the system can adopt. Figure 1a gives an example of such a configuration. The rdf gives the pair probability over several coordination shells. One may also specify SRO parameters involving clusters of 3, 4 and more sites (triplets, quadruplets and so on).

Figure 1b shows examples of particle clusters on a 2D square lattice, which are used to define the SRO-based correlations. These are clusters within a 4nn position cutoff. Red circles denote $A$ particles. The vacant species correspond to species $B$ here. The SRO parameter is defined as

$$z_c = \frac{N_c}{bN_A}, \quad 0 < N_A < N_t. \tag{1}$$

The subscript $c$ refers to a particular cluster in Figure 1b, $N_c$ is the cluster population in a configuration, $N_A$ is the $A$ population, $b$ is a geometric term such that the probability $z_c \in [0,1]$ and $N_t$ is the number of sites. See Ref. [17] for the value of $b$ in context of Figure 1. Note that $z_c$ is a property of a lattice configuration. In the canonical ensemble, the probability of a configuration $X$ depends on the composition, interactions $w$ and temperature $T$, i.e.,

$$p(X; N_A, N_t, w, T) = Q^{-1} \exp(-\beta E(X)). \tag{2}$$

Here $E$ is the energy, $\beta = (k_B T)^{-1}$, $k_B$ is the Boltzmann constant and $Q$ is a normalization factor. The average short-range order $\langle z_c \rangle$ is also a function of $N_A, N_t, w$ and $T$.

Of particular interest is the mapping: *lattice configuration* ⇌ *SRO parameters* at a single configuration level. We shall show that some SRO parameters are independent and the remaining are collinear. For example, clusters highlighted in Figure 1b (shown in red-font) are independent. To emphasize this aspect of dependence/independence we use a tilde/caret symbol. The caret symbol is used for the independent ones, e.g., $\hat{z}_{3nn}$, and the tilde symbol for the dependent ones, e.g., $\tilde{z}_{2nn}$. Thus, for any configuration $X$ we shall show that

$$\tilde{z} = \phi(x, \hat{z}). \tag{3}$$

Here $x = N_A/N_t$. Such fundamental relations arise from topological constraints on the lattice. An expression for $\phi$, if available, can be used to develop low-dimensional descriptions for the lattice dynamics and equilibrium. Here, $\langle \tilde{z} \rangle$ is expressed directly in terms $x$ and $\langle \hat{z} \rangle$, whereas, $\langle \hat{z} \rangle$ is dictated by $x, w$ and $T$. An analytical expression for $\phi$ cannot be derived because of the complex lattice topology. The problem becomes one of exploratory data analysis, where a large number of configurations for given $x, w$ and $T$ need to be examined to determine $\phi$, $\tilde{z}$ and $\hat{z}$.

The basis for this investigation lies in our previous studies (see Supplementary Material of Ref. [16]), where it was shown that in the solid solution forming Ag-Au alloy, the entire rdf can be generated from the 1nn pair probability. Thus, the second peak of the rdf, third peak and so on are related to the first one. $z_{1nn}$ describes for the local Ag-Au arrangement. On the other hand, in Ni-Pt, which forms ordered structure, the first three neighbor pair probabilities are required to generate the rdf. Thus, the full rdf contains redundant information. The hypothesis is that once 1nn, 2nn, 3nn pair probabilities are specified, there are only certain ways $A$ and $B$ particles can be arranged in subsequent neighbor positions while still satisfying the 1nn, 2nn, 3nn constraints. In effect, other SRO parameters depend on these constraints. The present study attempts to systematically find relationships of the form given by Equation (3).

In section 2, our methodology is described using a thermodynamic $A_xB_{1-x}$ system on a 2D square lattice. Results are discussed in section 3. Finally, conclusions are presented in section 4.

**2. Methodology**

The present study is of direct relevance to problems in catalysis, adsorption and surface science. Consider the situation where a single strongly binding species, such O, CO, halide or OH, is chemisorbed on a fcc(100) metal surface, such as, Pt, Pd or Cu (see Figure 1). The overlayer often exhibits interesting phase behavior, which arises out of adsorbate-adsorbate interactions. When the interactions are repulsive at the 1nn position, it causes the first neighbors to be absent. Therefore, we choose $z_{1nn} = 0$ in the present study (see Figure 1a). A lattice gas cluster expansion model of the form [18,19]

$$E(X) = \sum_c w_c N_c \tag{4}$$

is used here to capture the effect of adsorbate-adsorbate interactions. Here $E$ denotes the energy, $w_c$ is the cluster interaction for cluster $c$. See Section A4 in Supplementary Material regarding the origin of such an expression.

**Table 1. Models for adsorbate-adsorbate interactions (in eV) studied (see Figure 1b and Supplementary Material for more details). In all models, the 1nn interaction $w_{1nn} \to \infty$. Models 1-4 behave like model 0 at high temperatures.**

| Cluster interaction | Model 0 | Model 1 | Model 2 | Model 3 (Cl/Cu(100)) | Model 4 (Br/Cu(100)) |
|---|---|---|---|---|---|
| $w_{2nn}$ | 0 | -0.1 | 0.1 | 0.0167 | 0.0237 |
| $w_{3nn}$ | 0 | 0 | 0 | 0.0018 | -0.0148 |
| $w_{4nn}$ | 0 | 0 | 0 | 0.0002 | 0.002 |
| $w_I$ | 0 | 0 | 0 | 0.0163 | 0.0383 |
| $w_L$ | 0 | 0 | 0 | 0.0107 | 0.0243 |
| $w_V$ | 0 | 0 | 0 | 0.0091 | 0.0143 |
| $w_\Delta$ | 0 | 0 | 0 | -0.0042 | -0.0097 |
| $w_{Q1}$ | 0 | 0 | 0 | 0.0014 | 0.0077 |
| $w_{Q2}$ | 0 | 0 | 0 | 0.0004 | 0.0002 |
| $w_{Q3}$ | 0 | 0 | 0 | -0.0039 | -0.0051 |

Five different interaction models are studied. In all models, $w_{1NN}$ is set to infinity. Other interactions are listed in Table 1. Interactions strengths of up to $0.1\ eV$ are quite common. Model 1 involves attractive interactions that promote clustering at the 2nn position. Models 2-4 contain repulsive 2nn interactions. In models 3 and 4, the interactions involving all site clusters in Figure 1b are present. These interactions were obtained using density theory functional (DFT) calculations and correspond to the adsorbate interactions in Cl/Cu(100) and Br/Cu(100) system, respectively. The interested reader can refer to Ref. [20] for more details. Both the Cl and Br overlayers form an ordered c(2×2) structure in the high coverage limit. At high temperatures, models 1-4 behave like model 0.

Results are presented in the canonical ensemble, i.e., fixed $N_A$, $N_t$ and $T$. Suppose one partitions the lattice into smaller equal-sized domains. Consider a domain $d$ where $N_A^d$, $N_t^d$ and $N_c^d$ are the local particle, site and cluster populations. We require

$$N_A = \sum_d N_A^d, \tag{5}$$

$$N_t = \sum_d N_t^d, \tag{6}$$

and

$$N_c = \sum_d N_c^d. \tag{7}$$

Anticipating a scale-invariant relation between $\hat{z}$ and $\tilde{z}$ at both the full-lattice and domain level, the right-hand side of Equation (7) becomes

$$\tilde{N}_l = \sum_d \psi(\widehat{N}^d) \tag{8}$$

where $\psi$ is related to $\phi$; the former is used with cluster populations. For large domains where size effect is negligible, this relation can be satisfied for arbitrary domains only when $\psi$ is a linear function. Thus, at the domain level

$$\widetilde{N}_l^d = a_0 N_A^d + \sum_k a_k \widehat{N}_k^d. \tag{9}$$

$a_k$ is the associated coefficient. The point cluster is excluded in the sum. From Equation (9)

$$\tilde{z}_l^d = a_0 b_l^{-1} + \sum_k a_k b_l^{-1} b_k \hat{z}_k^d. \tag{10}$$

A similar expression is obtained at the lattice configuration level

$$\tilde{z}_l = a_0 b_l^{-1} + \sum_k a_k b_l^{-1} b_k \hat{z}_k. \tag{11}$$

A computational workflow is needed to uncover these underlying patterns, if they exist. One could sample cluster populations using MMC. For large lattices, the cluster population distribution is narrow and the sampled configurations may lack the diversity needed to properly identify $\tilde{z}$ and $\hat{z}$. Sampling of rare configurations in MMC is an issue. The approach used here involves small periodic lattices where all configurations are directly enumerated without MMC. This ensures that low probability configurations are included. A common problem in data analysis is that the conclusion reached is as good as the data. Such an issue does not arise here since the entire configurational space is sampled. The effect of lattice size can be also systematically probed. Once $\tilde{z}$ and $\hat{z}$ are identified, the coefficients in Equation (9)-(11) are determined for large lattices using MMC simulations.

**3. Results and discussion**

**3.1 Direct enumeration of lattice configurations**

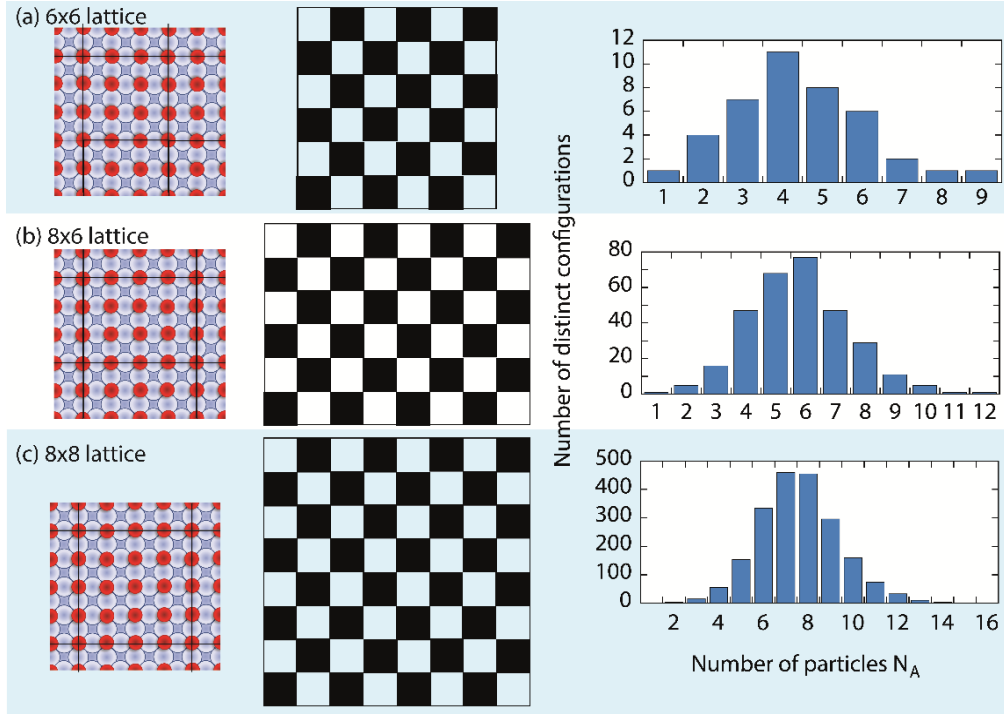

**Figure 2. Left: Periodic c(2×2) overlayer (red circles) on a substrate (blue). Middle: Corresponding $p \times q$ lattice. Right: Number of distinct configurations on the lattice. Rows: (a) $p = q = 6$, (b) $p = 8, q = 6$ and (c) $p = q = 8$.**

The left panels in Figure 2 show a c(2×2) overlayer. Black lines denote the boundaries of the periodic box. The corresponding lattice in middle panels contains black and white cells. The black cells are the top sites on fcc(100). These sites are ignored. The white (hollow sites) cells are occupied for the c(2×2) overlayer. A periodic $p \times q$ lattice contains $N_t = pq/2$ adsorption sites. $p$ and $q$ are chosen to be even integers for a commensurate overlayer to exist. Since 1nn pairs are absent, fewer configurations are possible, e.g., 473,862 out of $2^{N_t}$ =4.3 billion configurations for $N_t = 32$. A brute-force approach is pursued, wherein all configurations with $N_A \in [1, \frac{N_t}{2} - 1]$ are generated. From Equation (4), configurations having the same cluster population are associated with the same $E$. These configurations are grouped as a single configuration $X$. There are $\Omega_X$ such configurations. The multiplicity $\Omega_X$ can reach 1000 for $8 \times 8$ lattice (Figure 3). The number of unique/distinct configurations (unique in terms of cluster populations) with the $8 \times 8$ lattice is 2057. The right panels in Figure 2 show the breakup for the different lattices. In the canonical ensemble, the probability of a configuration $X$ is

$$p(X; N_A, N_t, w, T) = \frac{\Omega_X \exp(-\beta E(X))}{\sum_{X'} \Omega_{X'} \exp(-\beta E(X'))}. \tag{12}$$

The sum in the denominator is performed over all configurations with fixed $N_A$.

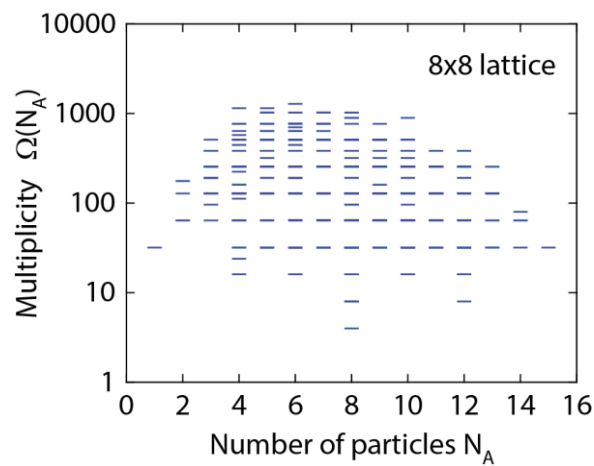

**Figure 3. Multiplicity for distinct lattice configurations on a 8x8 lattice where 1nn pairs are disallowed.**

Figure 4 shows the configurational space in terms of the cluster population $N_c$ plotted against $N_A$. Results for different interaction models are shown at 100 K. Recall that the same set of configurations is considered with all interaction models. The cluster population for each configuration is plotted in circles. The relative probability of a configuration $p_r(X)$ calculated with respect to the most likely configuration with same $N_A$ depends on the interaction and temperature. This aspect is highlighted in Figure 4 in terms of the size of the circle symbol, which is related to $\ln p_r(X)$. Low probability configurations appear as a dot. A general feature is that $N_c$ increases with $N_A$ indicating a positive correlation between the two variables. In fact, $N_A$ appears to capture most of the variation in $N_c$. The Supplementary Material contains plots for all clusters and different lattice sizes at temperatures of 100-700 K. At 700 K, the configuration probabilities for models 1-4 approach the ones for model 0. Here we mainly discuss results for 100 K where a greater effect of particle interactions is visible.

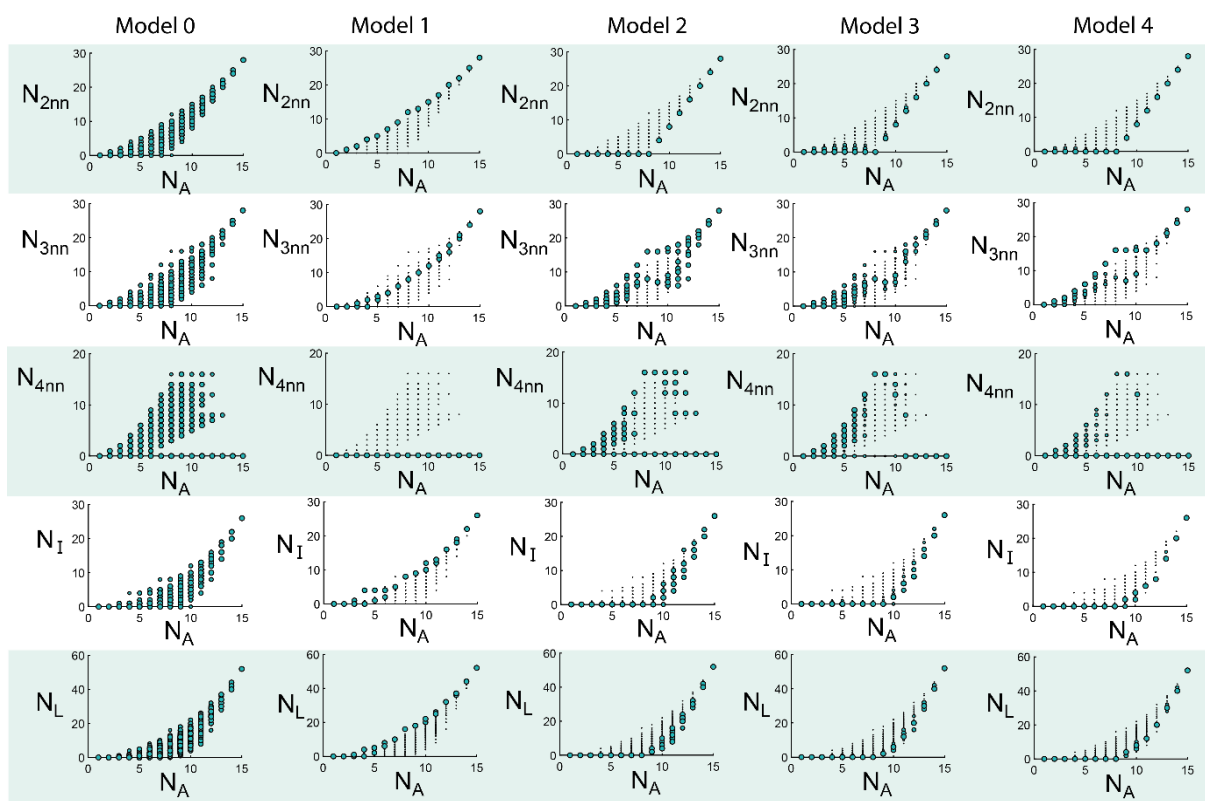

**Figure 4. Cluster population obtained with the different interaction models on an $8 \times 8$ lattice at $T = 100\ K$. See Table 1 for the interaction models.**

For model 0,

$$p(X; N_A, N_t, \beta w = 0) = \frac{\Omega_X}{\sum_{X'} \Omega_{X'}}. \quad (13)$$

A large vertical spread in the cluster population is observed especially for intermediate values of $N_A$ (see 2nn, 3nn, I and L clusters). At a single configuration level, $N_c$ cannot be uniquely determined from $N_A$ alone. The behavior changes in models 1-4. In the case of attractive 2nn interactions (model 1), configurations with large value of $N_{2nn}$ are preferred. Since other cluster interactions are absent, it is natural to expect a spread in other cluster populations analogous to model 1. However, this is not seen and other clusters also have characteristic populations. Certain aspects can be explained. For example, the presence of 4nn pairs prevents a 2nn pair from forming. Models 2-4 involve repulsive interactions at the 2nn position, therefore, smaller values of $N_{2nn}$ are preferred. The overall characteristics of the cluster populations in models 2-4 are similar. A spread in $N_{3nn}$ and $N_{4nn}$ is observed.

**3.2 Principal component analysis**

There are 11 $N$-particle clusters ($N = 1 - 4$) feasible with a cutoff of 4nn, such that 1nn pair are not involved. These are shown in Figure 1b. Assuming linear relations involving these 11 variables (equations (9)-(11)), then the principal component analysis (PCA) can be used to identify an orthonormal basis. PCA is a linear dimensionality reduction technique. The resulting directions, i.e., the principal components, provide a transformed coordinate system. Most of the variation in the data is captured with few principal components. The number of such directions provides the number of independent SRO parameters.

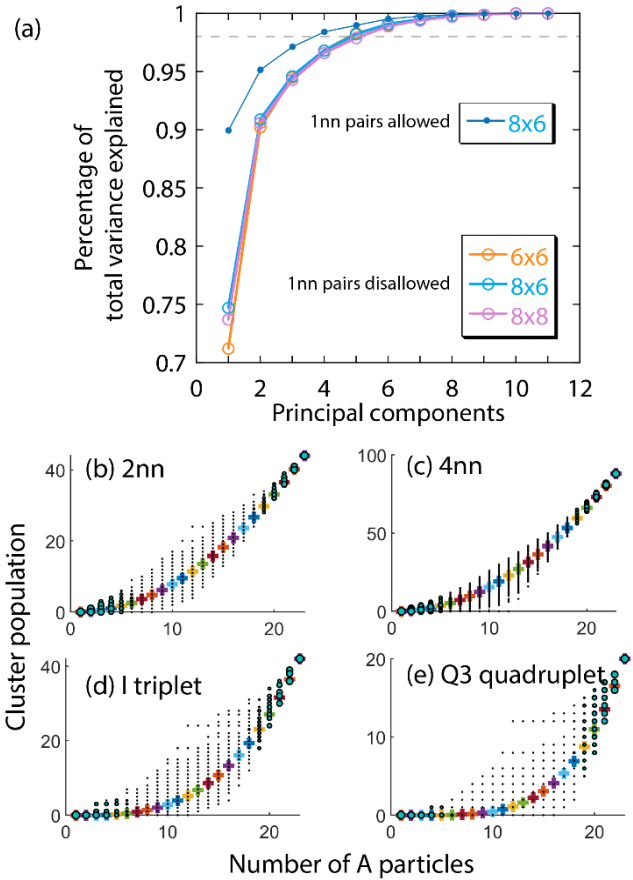

Figure 5. (a) Principal component analysis of cluster populations on a $p \times q$ lattice with model 0, i.e., particles disallowed at 1nn position. Results for random arrangement with particles allowed at 1nn position are also shown in panel (a) and the corresponding cluster populations in panels (b)-(e). Clusters in Figure 1 other than 1nn pair are analyzed.

Figure 5a shows the percentage variance explained by the top few principal components. Results are shown for three different lattices. The number of distinct configurations in the 6×6, 8×6 and 8×8 lattice is 40, 307 and 2057, respectively. The overall behavior for the three lattice sizes is similar. More than 98% of the total variance is captured using 5 principal components. The horizontal dashed line corresponds to 98%. The variation in the remaining six principle components is small. The species/point-cluster population $N_A$ is a natural choice as an independent variable. Unfortunately, since the principal components are linear combinations of the 11 cluster populations, they do not reveal which SRO parameters are independent, or whether an arbitrary choice can be made.

Before we proceed to determine the four independent SRO clusters, we discuss the origin of the reduced dimensionality. It is clear that disallowing particles in the 1nn position spatial introduces spatial correlations. To assess whether these spatial correlations are responsible for dimensionality reduction, consider the case when 1nn positions are allowed to be filled. Since there are no restrictions placed on the configurations, spatial correlations are completely absent. Figure 5(b)-(e) shows the cluster populations from such configurations. Note that the circle symbol size here is related to the absolute probability. $N_c$ has a wide distribution for intermediate value of $N_A$. Thus, the absolute probability of each configuration is small, and hence the small circle. The ensemble-averaged cluster population $\langle N_c \rangle$, which is shown in plus symbols. Analytical expression for $\langle N_c \rangle$ are available. For instance, $\langle N_{2nn} \rangle = \frac{2N_A(N_A-1)}{(N_t-1)}$, $\langle N_I \rangle = \frac{2N_A(N_A-1)(N_A-2)}{(N_t-1)(N_t-2)}$ and $\langle N_{Q3} \rangle = \frac{N_A(N_A-1)(N_A-2)(N_A-3)}{(N_t-1)(N_t-2)(N_t-3)}$. PCA of the cluster population reveals that the variation in the data can be captured by three principal components (see blue filled circles in Figure 5a labelled as 1nn pairs allowed). Once again the interpretation is that as more cluster populations are specified, the lattice topology imposes constraints on other cluster populations. This a fundamental property of the 2D square lattice.

**3.3 Identifying the independent SRO parameters**

As already witnessed in Figure 4, the high probability configurations in interaction models 1-4 can be different from model 0. The ordering behavior is dictated by interactions between the particles. Intuitively, one would expect the independent SRO parameters for model 2 to be different from model 0 and 1. Next, we discuss a procedure to analyze this aspect.

Consider configurations having a relative probability (calculated with respect to the most likely configuration with the same $N_A$) greater than $10^{-3}$. The collection of such high-probability configurations is denoted as $V$. First, we need to define $\hat{z}$ for the given interactions and temperature. For this purpose, different combinations of SRO parameters are chosen as a putative $\hat{z}$. This is done by exploring tuples of $m$ different SRO parameters. To achieve dimensionality reduction, $m$ should be less than 10. Ideally, small values of $m$ are preferred for reasons of compactness. When $m = 1$, the tuples of interest are $\{z_{2nn}\}$, $\{z_{3nn}\}$, …, $\{z_{Q_3}\}$. For $m = 2$, the tuples $\{z_{2nn}, z_{3nn}\}$, $\{z_{2nn}, z_{4nn}\}$, … are explored, and so on.

**Table 2. Number of SRO parameters combinations explored in order to identify $\hat{z}$. The species population is also specified by default and is not regarded as an SRO parameter.**

| Number of SRO parameters | Combinations |
|---|---|
| 1 | 10 |
| 2 | 45 |
| 3 | 120 |
| 4 | 210 |
| 5 | 252 |
| 6 | 210 |
| 7 | 120 |
| 8 | 45 |
| 9 | 10 |

Out of the 1022 combinations possible (see Table 2), the quality of $\hat{z}$ as an independent parameter is assessed in terms of two quantities, the mean squared error (MSE) and the entropy, as discussed next. These quantities measure the variability in the remaining SRO parameters. A low value of MSE and entropy (preferably close to zero) is desirable. Recall that Figure 3 exhibits a large multiplicity (>1000) for intermediate values $N_A$. A particular combination of $N_A$ and $\hat{z}$ will be associated with smaller multiplicity.

For a configuration $X \in V$, both $\hat{z}$ and $\tilde{z}$ are known to us. The particle and the respective cluster populations are $N_A(X)$, $\widehat{N}(X)$ and $\widetilde{N}(X)$. Suppose a configuration $X^r$ is to be reconstructed with the constraint $N_A(X^r) = N_A(X)$ and $\widehat{N}(X^r) = \widehat{N}(X)$. The superscript $r$ denotes the reconstructed configuration/quantity. $X^r$ is picked from the configurations available for model 0 with a probability that is analogous to Equation (13) while incorporating the constraint $\widehat{N}(X^r) = \widehat{N}(X)$. We use the notation $p(X^r; N_A, \widehat{N})$ to denote this probability. The corresponding cluster population is $\widetilde{N}^r(X^r; N_A, \widehat{N})$. Next, anticipating a large variability in $\widetilde{N}^r$, consider the working collection of reconstructed configurations $W(N_A, \widehat{N})$ with the specified constraint, such that $X^r \in W$. The average cluster population in $W$ is

$$\langle \widetilde{N}^r(N_A, \widehat{N}) \rangle_W = \sum_{X^r \in W(N_A, \widehat{N})} p(X^r; N_A, \widehat{N}) \widetilde{N}^r(X^r; N_A, \widehat{N}). \qquad (14)$$

The squared error for the average cluster population with respect to $X \in V$ using the putative $\hat{z}$ is

$$SE(X, \hat{z}) = \left\| \langle \widetilde{N}^r(N_A, \widehat{N}) \rangle_W - \widetilde{N}(X) \right\|^2. \qquad (15)$$

The mean-squared error for $V$ is defined as

$$MSE(V, \hat{z}) = \frac{1}{\dim(V)} \sum_{X \in V} SE(X, \hat{z}) \qquad (16)$$

where $\dim(V)$ is the size of $V$, i.e., the number of high-probability configurations. In equation (16), notice that the number of particles $N_A$ in $V$ is allowed to be variable.

The information entropy for the configuration $X$ using the putative $\hat{z}$ is defined as

$$S(X, \hat{z}) = - \sum_{X^r \in W(N_A, \widehat{N})} p(X^r; N_A, \widehat{N}) \ln p(X^r; N_A, \widehat{N}). \qquad (17)$$

This equation measures the spread in $p(X^r; N_A, \widehat{N})$. Finally, the total information entropy is

$$S(V,\hat{z}) = \sum_{X \in V} S(X,\hat{z}). \tag{18}$$

Notice that all configurations $X \in V$ contribute with the same weight to the MSE and information entropy. From Figure 4, configurations with extreme values of $N_A$ tend to have low variability in the cluster populations. Therefore, $SE(X,\hat{z})$ and $S(X,\hat{z})$ should be low for these contributions. On the other hand, intermediate values of $N_A$ typically result in greater variability and are expected to contribute more. Next, we analyze the behavior of $MSE(V,\hat{z})$ and $S(V,\hat{z})$.

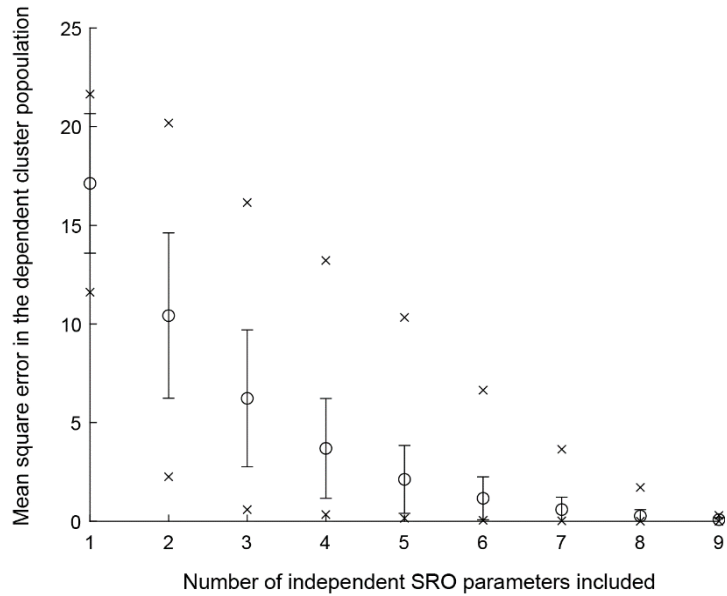

**Figure 6. Mean squared error as more SRO parameters are treated as being independent. Results are shown for model 3 (Cl/Cu(100)) on $8 \times 8$ lattice at a temperature of 100 K.**

The general trends for the MSE and entropy are qualitatively similar for all interaction models, lattices and temperatures considered here. The complete analysis can be found in the Supplementary Material. Here we focus mainly on the results for model 3 with a 8×8 lattice. Recall that in model 3 all clusters contribute to the energy of the configuration (see Table 1).

Figure 6 shows the MSE when $m$ SRO parameters are specified. The horizontal axis does not exactly specify which SRO parameters are involved in $\hat{z}$. However, the minimum and maximum values of MSE are shown by the cross symbol. For small $m$, there is a large gap between the maximum and minimum highlighting an important observation that certain SRO parameters are better descriptors for the local arrangement. This aspect was not evident from the principal component analysis. The average MSE for a given $m$ is shown in the open circle and the corresponding standard deviation is shown as bar. The MSE decreases with $m$. In Figure 6, for $m = 4$ the MSE is least (0.34) when $\hat{z} \equiv \{z_{2nn}, z_{3nn}, z_{4nn}, z_\Delta\}$. The MSE can be further lowered by specifying more SRO parameters. For instance, $\{z_{2nn}, z_{3nn}, z_{4nn}, z_V, z_\Delta\}$ has an MSE of 0.22 (see Supplementary Material).

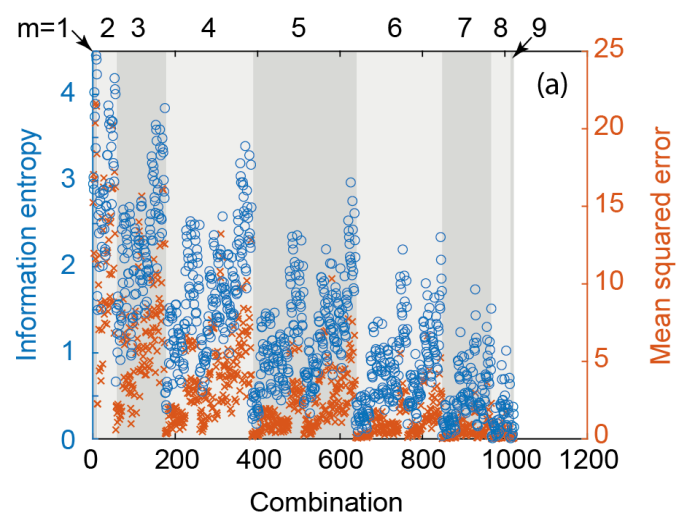

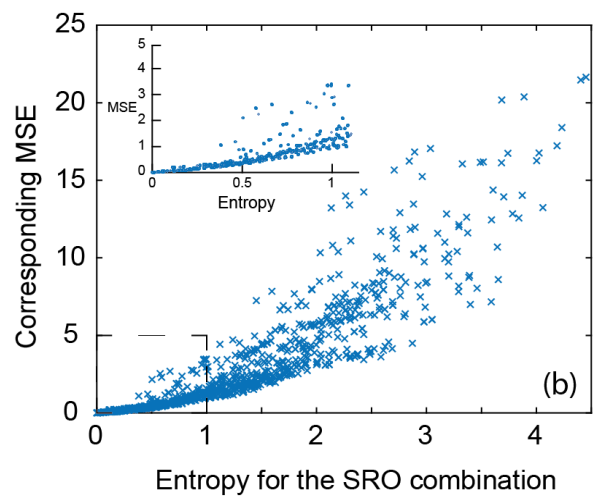

**Figure 7. Entropy and mean squared error for 1022 different combinations of SRO parameters. Conditions are identical to Figure 6. Inset in panel (b) shows a zoomed version of the dashed rectangle.**

Figure 7a, which shows the MSE and entropy for the SRO combinations/tuples, provides some more insights. $m$ increases to the right along the horizontal axis. For this reason, the gray shaded bands are shown. The thickness of the band is related to Table 2. Suppose only one SRO parameter is specified, then the resulting entropy (open circles) and the MSE (cross) is high. The spread in $p(X^r; N_A, \widehat{N})$ measured in terms of the entropy is significant. The MSE can be as large as 20 suggesting that it is a more sensitive quantity for assessing the optimal SRO combination. As seen in Figure 6, specifying additional SRO parameters lowers both the entropy and MSE. Figure 7b shows the MSE plotted against the information entropy. Since a large spread in $p(X^r; N_A, \widehat{N})$ would also result in a large MSE, the MSE is directly related to the information entropy at least in the limit where $S \to 0$ (see inset in Figure 7b).

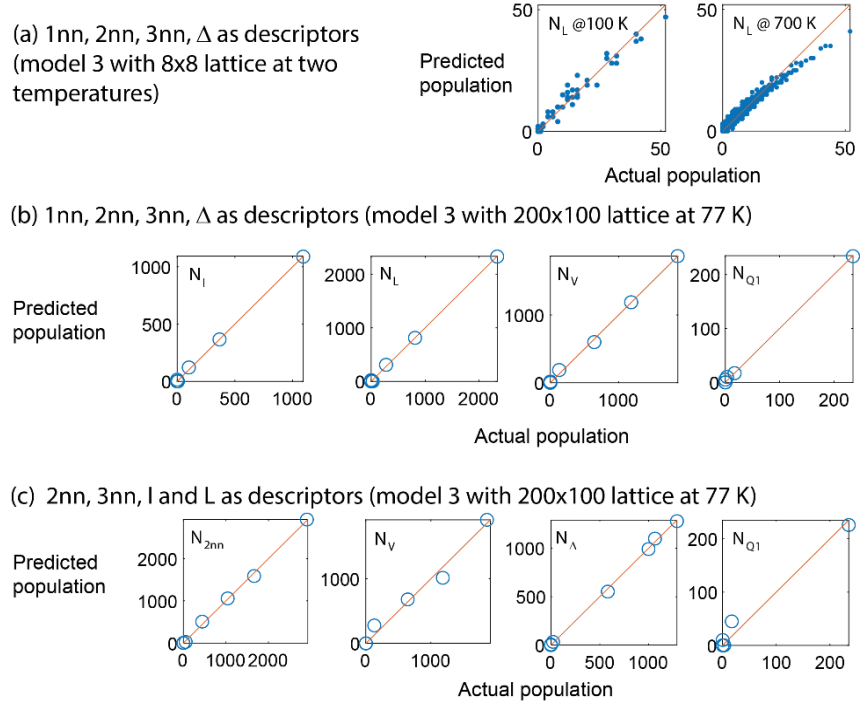

**Figure 8. Linear models derived for populations of dependent clusters in terms of the populations of independent ones.**

Note that the analysis of Figures 6 and 7 only assesses the quality of the assumed $\hat{z}$. It does not assume any linear relationship. However, if a linear model of the form given by Equations (9)-(11) were to be written, the coefficients $a_k$ can be determined via linear regression using the cluster population data in the high-probability configurations. Figure 8a shows one result from such an exercise. The parity plot gives the predicted-versus-actual population for cluster L at 100 K. At 700 K, the combination $\hat{z} \equiv \{z_{3nn}, z_{4nn}, z_I, z_L\}$ gives the least MSE. Typically, at any given temperature other SRO combinations may also result in low MSE and entropy, and these can be selected as descriptors. Accordingly, a dynamic/thermodynamic model can be formulated (see Section 4). SRO combinations with high MSE and entropy should be avoided.

Extending the analysis to model 0, such that all configurations are included in $V$, MSE and information entropy are determined for different SRO parameter combinations. The results are provided in Section B of the Supplementary Material. Similar to behavior for model 3 at 700 K, the SRO parameters $\{z_{3nn}, z_{4nn}, z_I, z_L\}$ has the lowest MSE. At high temperatures, the local arrangements in other models 1, 2 and 4 are also best described by the same SRO parameters.

The data analysis supports the hypothesis given in the introduction section that spatial correlations introduce relationship between different SRO parameters (at least in context of models 0-4). The approach discussed so far may not be as feasible for much larger lattice sizes due to the sheer number of configurations possible. Nonetheless, our understanding can be extended to this situation. Consider a small region of the lattice for which the local $\hat{z}$ has been specified. This provides adequate enough information about the local ordering behavior. Because the same definition of $\hat{z}$ can be used for all local compositions, it becomes possible to write equations (9)-(11).

The coefficients in equations (9)-(11) can be obtained for large lattices by using an importance sampling approach. Here we perform grand-canonical Monte Carlo simulations at different Cl gas phase chemical potentials and a temperature of 77 K. The simulations are analogous to the ones reported in Ref. [20] except that for periodic lattice has a size $200 \times 100$. One equilibrium configuration is taken from each chemical potential and the cluster populations are noted. Using the low temperature descriptor from the 8×8 lattice, i.e., the SRO parameter combination $\{z_{2nn}, z_{3nn}, z_{4nn}, z_\Delta\}$, a linear model is fitted and the coefficients $a_k$ in equations (9)-(11) are obtained:

$$N_I = 0.775 N_{2nn} + 0.063 N_{3nn} - 0.882 N_{4nn} + 1.33\, N_\Delta; \quad R^2 = 1, \qquad (19)$$

$$N_L = 1.62N_{2nn} + 0.13N_{3nn} - 1.88N_{4nn} + 2.93N_\Delta; \quad R^2 = 1,$$

$$N_V = 0.394N_{2nn} - 0.154N_{3nn} + 1.58N_{4nn} - 3.34N_\Delta; \quad R^2 = 1,$$

$$N_{Q1} = 0.277N_{2nn} + 0.052N_{3nn} - 0.68N_{4nn} + 1.26N_\Delta; \quad R^2 = 0.99.$$

Equation (19) provides a list of topological constraints for the 2D square lattice. Figure 8b shows a parity plot confirming the quality of the model. The Pearson correlation coefficient $R^2 \approx 1$ for all models constructed. A lower-quality model is obtained when the SRO combination $\{z_{2nn}, z_{3nn}, z_{4nn}\}$ is used as the low temperature as one of the models generated has $R^2 = 0.93$. Similarly, suppose the high temperature descriptor from the 8×8 lattice, i.e., $\{z_{3nn}, z_{4nn}, z_I, z_L\}$, is used, the performance is suboptimal.

**4. Implications**

The species balance equation is a fundamental conservation equation used to understand physical systems. A modeling framework is required that can incorporate particle interactions, topological constraints and ordering behavior as building blocks to understand the macroscopic emergent behavior. Recall that the relative probability of $10^{-3}$ used earlier. This ensures that rare configurations have been included in the analysis and away from equilibrium situations can be handled. The species balance equation is written as

$$\frac{dN_A}{dt} = f(N_A, \widehat{N}, \widetilde{N}). \tag{20}$$

The idea behind Equation (20) is that the rates of elementary processes can be expressed in terms of the local chemical environment and the interactions. The cluster populations are bound to change in time. Therefore, Equation (20) is augmented with an evolution equation for cluster population

$$\frac{d\widehat{N}}{dt} = g(N_A, \widehat{N}, \widetilde{N}). \tag{21}$$

Additionally, the lattice topological constraints are included, i.e., $\widetilde{N}$ and $\widehat{N}$ are related through equations (8)-(11). These equations embed the atomistic structural details into the continuum model (Equations (20) and (21)). Note that cluster populations are non-conserved. The right-hand side contains growth rate for the species and cluster populations, which typically contain rate parameters for elementary physical/chemical processes, such as adsorption, desorption, diffusion, reaction, etc.

Equation (21) describes the SRO evolution (see equation (1) for relation between $\widehat{N}$ and $\hat{z}$). Such a non-equilibrium modeling framework has been recently proposed in Ref. [21]. The main limitation of the earlier work was that $\hat{z}$ was chosen using intuition. Now a systematic procedure is available. Although Equations (20) and (21) are continuum models, a 2D or 3D atomistic configuration can be reconstructed using the reverse Monte Carlo method [22,23].

A thermodynamic model can be arrived at simply by setting the time-derivative in Equations (20) and (21) to zero. The equilibrium populations $N_A^{eq}$ and $\widehat{N}^{eq}$ are determined using a numerical technique solves for the roots of the right-hand side of Equations (20) and (21). The equilibrium phase behavior and properties are obtained from $N_A^{eq}$ and $\widehat{N}^{eq}$ [24]. Such an approach has been discussed in Ref. [20].

The present study also enables the development of a hierarchy of lattice models by deciding which SRO parameters to include. From our earlier discussion, a high-accuracy model should contain at least four SRO parameters, e.g., $\widehat{N} = \{N_{2nn}, N_{3nn}, N_{4nn}, N_\Delta\}$ at low temperatures. Models of higher accuracy can be created by including those SRO parameters that add most information. Figure 7 provides the guidelines for the selection of optimal SRO parameters. A model treating all SRO parameters as independent is unnecessary and certain combinations of cluster populations may not be even physically realizable.

## 5. Conclusions

Short-range order (SRO) parameters are used extensively in literature to quantify spatial correlations in condensed matter. SRO parameters have traditionally served as building blocks to piece together the many-particle distribution and to help understand the emergent macroscopic behavior. This paper discusses the nature of SRO parameters in lattice-based interacting particle systems. It is well known that the ordering behavior, e.g., the radial distribution function, arises from particle interactions. However, the fact that topological constraints imposed by the lattice may have an effect on the SRO parameters is somewhat underappreciated.

A fundamental property of lattice systems is that only certain SRO parameters (these could be combinations of pairs, triplets and so on) are truly independent and the remaining (other pairs, triplets, etc.) are intricately linked to the independent ones. For the 2D square lattice, four SRO parameters act as descriptors for the local particle arrangements. Other SRO parameters can be determined from these four. This property is observed at a single lattice configuration level itself and it arises because of topological constraints in the underlying lattice. Remarkably, the property holds even when spatial correlations are absent.

A procedure to systematically identify the independent SRO parameters and determine the topological constraints is introduced. A descriptor having relatively few SRO parameters and low information entropy/mean squared error is optimal. The same descriptor can successfully explain the complexity in a variety of lattice configurations, interactions and temperature. However, if the combination of SRO parameters with the lowest entropy/mean squared error is sought, then it is possible that the combination may change with temperature and particle interactions. For the different interaction models investigated here, a linear relationship between the SRO parameters is discovered.

An important question arises about how these topological constraints can be incorporated into physical models. To answer this question, a reduced-dimensionality (dynamical and thermodynamic) modeling framework has been proposed that is written in terms of the independent SRO parameters. The work provides the essential building blocks needed to study complex systems where the local atomic arrangements lead to emergent macroscopic behavior.

## Acknowledgements

AC acknowledges support from Science and Engineering Research Board Grant No. CRG/2022/008058 and National Supercomputing Mission DST/NSM/R&D_HPC_Applications/2021/02. Density functional theory and grand canonical Monte Carlo simulations were performed by Bibek Dash and Suhail Haque, respectively.

## Supplementary Material

Supplementary Material has been provided.

## Data Availability Statement

The data that supports the findings of this study are available within the article and its supplementary material.

[10] T. O. Drews, R. D. Braatz, and R. C. Alkire, *Parameter Sensitivity Analysis of Monte Carlo Simulations of Copper Electrodeposition with Multiple Additives*, J. Electrochem. Soc. **150**, C807 (2003).

[11] D. Frenkel and B. Smit, *Understanding Molecular Simulation: From Algorithms to Applications* (Academic Press, New York, 1996).

[12] D. P. Landau and K. Binder, *A Guide to Monte Carlo Simulations in Statistical Physics* (Cambridge University Press, Cambridge, 2000).

[13] M. P. Allen and D. J. Tildesley, *Computer Simulation of Liquids* (Oxford Science Publications, Oxford, 1989).

[14] G. Agrahari and A. Chatterjee, *Speed-up of Monte Carlo Simulations by Preparing Starting off-Lattice Structures That Are Close to Equilibrium*, J. Chem. Phys. **152**, 44102 (2020).

[15] J. P. HANSEN and I. R. McDONALD, *Theory of Simple Liquids*, 3rd ed. (Academic Press, Burlington, 2006).

[16] G. Agrahari and A. Chatterjee, *Thermodynamic Calculations Using Reverse Monte Carlo: Convergence Aspects, Sources of Error and Guidelines for Improving Accuracy*, Mol. Simul. **48**, 1143 (2022).

[17] S. Haque and A. Chatterjee, *Thermodynamic Calculations Using Reverse Monte Carlo: Simultaneously Tuning Multiple Short-Range Order Parameters for 2D Lattice Adsorption Problem*, J. Chem. Phys. **159**, 104106 (2023).

[18] G. Collinge, K. Groden, C. Stampfl, and J.-S. McEwen, *Formulation of Multicomponent Lattice Gas Model Cluster Expansions Parameterized on Ab Initio Data: An Introduction
27